\documentclass[journal]{IEEEtran}
\usepackage{amssymb}
\usepackage[figuresright]{rotating}
\usepackage[tableposition=top]{caption}
\usepackage{amsmath,subfigure,epsfig,bbding}
\usepackage{epstopdf}
\usepackage{multicol}
\usepackage{booktabs}
\usepackage{multirow}
\usepackage{threeparttable}
\usepackage{color}
\usepackage{makecell}
\usepackage{amsmath}
\usepackage[linesnumbered,ruled,vlined]{algorithm2e}
\usepackage{algpseudocode}
\usepackage{amsmath}
\usepackage{graphicx}
\usepackage{bm}
\usepackage{array}
\newcommand{\PreserveBackslash}[1]{\let\temp=\\#1\let\\=\temp}
\newcolumntype{C}[1]{>{\PreserveBackslash\centering}p{#1}}
\newcolumntype{R}[1]{>{\PreserveBackslash\raggedleft}p{#1}}
\newcolumntype{L}[1]{>{\PreserveBackslash\raggedright}p{#1}}

\begin{document}

\title{Fast Reinforcement Learning\\
for Anti-jamming Communications}
\author{Pei-Gen~Ye, Yuan-Gen~Wang,~\IEEEmembership{Senior Member,~IEEE,} \\Jin Li,~\IEEEmembership{Senior Member,~IEEE,} and Liang Xiao,~\IEEEmembership{Senior Member,~IEEE}
\thanks{P.-G.~Ye, Y.-G. Wang, and J. Li are with the School of Computer Science and Cyber Engineering, Guangzhou University, Guangzhou 510006, China (E-mail: ypgmhxy@126.com, \{wangyg, lijin\}@gzhu.edu.cn).

L. Xiao is with the Department of Communication Engineering, Xiamen University, Xiamen 361005, China (E-mail: lxiao@xmu.edu.cn).}}
\maketitle

\begin{abstract}
This letter presents a fast reinforcement learning algorithm for anti-jamming communications which chooses previous action with probability $\tau$ and applies $\epsilon$-greedy with probability $(1-\tau)$. A dynamic threshold based on the average value of previous several actions is designed and probability $\tau$ is formulated as a Gaussian-like function to guide the wireless devices. As a concrete example, the proposed algorithm is implemented in a wireless communication system against multiple jammers. Experimental results demonstrate that the proposed algorithm exceeds Q-learing, deep Q-networks (DQN), double DQN (DDQN), and prioritized experience reply based DDQN (PDDQN), in terms of signal-to-interference-plus-noise ratio and convergence rate.
\end{abstract}


\section{INTRODUCTION}
\IEEEPARstart{W}{ireless} communications are highly vulnerable to jamming attacks due to the open and sharing nature of wireless medium \cite{Mukherjee2014}. In general, attackers jam the ongoing transmissions via injecting malicious signal to wireless channel in use. To deal with the jamming attacks, the frequency hopping was proposed by selecting ``good'' frequencies in an ad-hoc way and avoiding the jammed frequency \cite{Lance1997, Wang2011}. But, it is no guidance for the random frequency hopping techniques to select a channel that is not blocked. Soon afterwards, Wang \emph{et al.} \cite{Wang2012} proposed an uncoordinated frequency hopping method by using the online learning theory to mitigate this problem. However, this conventional online learning method achieves only asymptotic optimum.

Recently, the success of reinforcement learning (RL) in decision-making problem attracts researchers to apply Q-learning (QL) to anti-jamming wireless communications. For instance, Xiao \emph{et al.} \cite{Xiao2015, Xiao2018d} employ QL to choose an appropriate transmission power. In \cite{Gwon2013, Slimeni2015}, QL is applied to select optimal frequency hopping channel. However, as the dimension of the actions increases, the Q-table of QL will become too large to quickly derive the optimal policy. Subsequently, deep Q-network (DQN) \cite{Mnih2015} was proposed to solve such a high-dimensional optimization problem. Different from QL, the Q value of DQN is not calculated by the state-value function, but learned by the neural networks such as convolutional neural network (CNN) and recurrent neural network (RNN). For example, in order to accelerate the learning speed and enhance the anti-jamming communication performance, Han \emph{et al.} \cite{Han2017} apply DQN to choose the optimal communication channel, and Liu \emph{et al.} \cite{Liu2018} combine DQN and spectrum waterfall to improve the ability of exploring the unknown environment. With the rapid development of RL, double DQN (DDQN) \cite{Hasselt2016} has been proposed, which combines two deep networks to avoid the dependency between the Q value calculation and network update. During the backpropagation of DDQN, only one network used for action selection is updated and its parameters are directly copied to the other network with a constant frequency. Hence, to some extent, DDQN mitigates the overestimation problem. Furthermore, prioritized experience reply based DQN (PDQN) \cite{Schaul2015} was developed to optimize the experience replay of DQN. Interestingly, researchers promptly set out to combine DDQN and PDQN together (a.k.a. PDDQN) to improve the convergence rate and avoid the overestimation problem, such as in game design \cite{Hester2018}. Besides, the action repetition theory \cite{Aravind2017} was proposed to further enhance the convergence rate of DQN, whereas optimizing the action repetition rate becomes difficult. It is important and challenging to apply PDDQN to solve the optimization problem with the anti-jamming wireless communications.

In this letter, we present a fast reinforcement learning algorithm called $(\tau, \epsilon)$-greedy. The key idea of $(\tau, \epsilon)$-greedy is to preserve previous action with probability $\tau$ and apply $\epsilon$-greedy with probability $1-\tau$. Specifically, probability $\tau$ is formulated as a Gaussian-like function such that the greater value the previous action has, the larger value $\tau$ takes. By doing so, the optimal action can be selected with a higher probability and thus the learning process is significantly accelerated. As a concrete example, the proposed algorithm is implemented in a wireless communication system against multiple jammers. Experimental results show the superiority of the proposed algorithm over existing value-based RLs including QL, DQN, DDQN, and PDDQN. Especially, the $(\tau, \epsilon)$-greedy based PDDQN achieves the best performance in terms of signal-to-interference-plus-noise ratio (SINR) and convergence rate.

The rest of this letter is organized as follows. In Section II, we introduce an anti-jamming wireless communication model. The proposed algorithm is described in detail in Section III. Section IV provides experimental results and conclusions are drawn in Section V.

\section{SYSTEM MODEL}
As depicted in Fig. 1, we consider a wireless communication system where a sender transmits data to the receiver with a power $P_{S}(s)$, while there are $J$ jammers who can simultaneously launch jamming attacks with $L$ different power levels on channels. The $L$ jamming power levels are denoted as $P_{J}(l)\in\{P_{J}(1),P_{J}(2),...,P_{J}(L)\}$. In order to resist this complicated attack, the sender dynamically chooses an appropriate $P_{S}(s)$ from $S$ different power levels $\{P_{S}(1),P_{S}(2),...,P_{S}(S)\}$, which in general is shown to be superior to the use of constant power under the constraint of the same average power \cite{Xiao2018a}. Note that both the sender and jammers share the same $N$ frequency channels. At time slot $k$, the channels chosen by the sender and $J$ jammers are respectively denoted by $x^{(k)}$ and $\{y_{1}^{(k)},y_{2}^{(k)},...,y_{J}^{(k)}\}$. $h_{s}$ and $h_{j}$ denote the channel power gains from the sender and the $j$th jammer ($j\in\{1,...,J\}$) to the receiver, respectively.
\begin{figure}
\centering
{\label{SS1}
\includegraphics[width=3.4in]{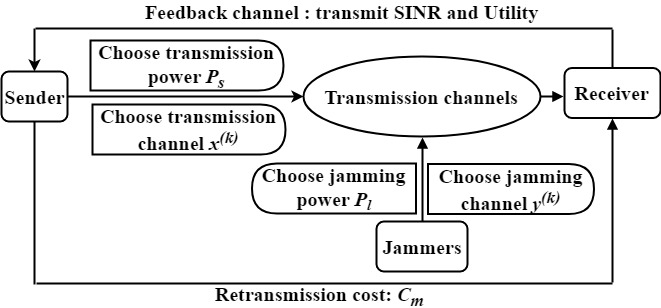}}
\caption{Anti-jamming wireless communication system}
\end{figure}

In the above communication system, after the receiver gets the signal at time slot $k$, the SINR is calculated by Eq. (\ref{SINR}) and returned to the sender in the feedback channel.
\begin{equation}
\textrm{SINR}= \frac{P_{S}(s)h_{s}}{\beta + \sum_{j=1}^{J}P_{J}^{j}(l)h_{j}f(x^{(k)}=y_{j}^{(k)})},
\label{SINR}
\end{equation}
where $\beta$ is the receiver noise power, $P_{J}^{j}(l)$ denotes the jamming power chosen by the $j$th jammer, and $f(\xi)$ is an indicator function that equals 1 if $\xi$ is true and 0 otherwise. If the receiver did not get the signal at time slot $k$, the transmission channel is considered to be blocked by the jammers. In this case, the receiver informs the sender about the channel state information through the feedback channel. Then the sender retransmits the signal. The additional cost caused by the retransmission is denoted as $C_{m}$. For simplicity, we assume that the feedback channel cannot be attacked and the transmission channel is considered to be blocked if $P_{J}(l)$ takes the maximum value $P_{J}(L)$. Besides, the cost of the unit transmit power is denoted as $C_{s}$. In \cite{Xiao2018b}, the authors formulated the utility function by considering the transit cost is proportional to the transmit power. Based on the SINR, retransmission cost $C_{m}$, and transit cost, in this letter the utility $u_{s}^{(k)}$ at time slot $k$ is defined by
\begin{equation}
u_{s}^{(k)}= \textrm{SINR}-C_{m}f(P_{J}^{j}(l)= P_{J}(L))f(x^{(k)}= y_{j}^{(k)})-C_{s}P_{s}.
\label{utility}
\end{equation}
It is known that the anti-jamming wireless communications aim to maximize the communication performance while at the same time save energy as much as possible. By referring to Eq. (\ref{utility}), our designed utility function has the ability of making a good tradeoff between the cost and communication performance.

\section{PROPOSED ALGORITHM}
The proposed algorithm consists of three modules: $(\tau, \epsilon)$-greedy action policy, double deep network structure, and prioritized experience replay, as shown in Fig. 2. The system state at time slot $k$ is denoted as $s^{(k)}=\textrm{SINR}^{(k-1)}$, where $\textrm{SINR}^{(k-1)}$ is the SINR at time slot $k-1$. Based on the current state $s^{(k)}$, the sender selects an action containing the selected transmission channel $x^{(k)}$ and power level $P^{(k)}_{S}(s)$ (denoted as $\textbf{a}^{(k)}=[x^{(k)},P^{(k)}_{S}(s)]$). After the action is adopted, the sender receives a reward (denoted as $u^{(k)}_{s}$). In the following, we describe each module in detail.

\subsection{$(\tau, \epsilon)$-greedy Action Policy}
In RL, the Q value is updated by a Q-function, which is written by
\begin{equation}
Q(s^{(k)}, \textbf{a}^{(k)})=E[u_{s}^{(k)}+\gamma \max \limits_{\textbf{a}'\in \mathcal{A}}Q(s^{(k+1)},\textbf{a}')\mid s^{(k)},\textbf{a}^{(k)}],
\label{qvalue}
\end{equation}
where $s^{(k+1)}$ denotes the next state provided the sender takes action $\textbf{a}^{(k)}$ at state $s^{(k)}$, $\mathcal{A}$ denotes a set of actions that can be chosen at state $s^{(k+1)}$, and $\gamma$ denotes the discount factor which represents the uncertainty of the sender about the future rewards. To our best knowledge, most existing value-based RL methods apply $\epsilon$-greedy to action selection policy. In conventional $\epsilon$-greedy, agents select the action with the maximum Q value with a high probability and randomly select an action with a very low probability $\frac{\epsilon}{|\mathcal{A}|}$ where $|\mathcal{A}|$ denotes the cardinality of set $\mathcal{A}$. However, we find that agents still need to calculate the Q value at the next time slot even if they have already chosen an action which can greatly improve the utility at the current time slot. Moreover, we consider that during the learning process, the value of $\epsilon$ should be gradually decreased to ensure that agents have a higher probability to explore the possible optimal actions at the beginning and then the Q function can converge quickly in the end.

Based on $\epsilon$-greedy, we propose to add a parameter $\tau$ to represent the probability of the current action $\textbf{a}^{(k)}$ that is directly selected at the next time slot without calculating the Q value, called $(\tau, \epsilon)$-greedy action policy in this letter. Thus, there will be three possible ways for the sender to select the action at the current state. That is the sender may directly adopt previous action with probability $\tau$, randomly select any action in $\mathcal{A}$ with probability $\frac{\epsilon}{|\mathcal{A}|}$, and take the action which has the maximum Q value with probability $(1-\tau-\epsilon)$. The proposed $(\tau, \epsilon)$-greedy is formulated as
\begin{equation}
\pi(\textbf{a}^{(k)}|s^{(k)})=
\begin{cases}
\textbf{a}^{(k-1)}&{\textrm{with }\tau}\\
\textbf{a}_{r}&{\textrm{with }\frac{\epsilon}{|\mathcal{A}|}}\\
\textrm{arg}\max \limits_{\textbf{a}'\in \mathcal{A}}Q(s^{(k)},\textbf{a}')&{\textrm{with } (1-\tau-\epsilon)}
\end{cases},
\label{policy}
\end{equation}
where $\pi$ and $\textbf{a}_{r}$ denote the action policy and a random action, respectively. Let $\overline{u}_{s}^{(k-1)}$ denote the average utility of previous $T$ time slots that can be computed by $\overline{u}_{s}^{(k-1)}= \frac{1}{T}\sum_{i=1}^{T}u_{s}^{(k-i)}$. According to $(\tau, \epsilon)$-greedy, the sender will directly take the action $\textbf{a}^{(k)}$ with probability $\tau$ at the next time slot ($k+1$). In order to improve the learning speed, we hope to preserve previous action which greatly contributes to the system. Therefore, we employ a difference between $u_{s}^{(k)}$ and $\overline{u}_{s}^{(k-1)}$ to measure how much the contribution of action $\textbf{a}^{(k)}$ is. It is reasonable that the probability $\tau$ of adopting $\textbf{a}^{(k)}$ at the next time slot should increase as the difference $(u_{s}^{(k)}-\overline{u}_{s}^{(k-1)})$ increases, vice versa. Based on this consideration, we design a Gaussian-like function to compute the value of $\tau$ as follows.
\begin{equation}
\tau =
\begin{cases}
1-\frac{1}{\sqrt{2\pi}\sigma_{1} }\textrm{exp}(\frac{(u_{s}^{(k)}-\overline{u}_{s}^{(k-1)})^{2}}{2\sigma_{1}^{2}}) &{u_{s}^{(k)}>\overline {u}_{s}^{(k-1)}}\\
\frac{1}{\sqrt{2\pi}\sigma_{2} }\textrm{exp}(\frac{(u_{s}^{(k)}-\overline{u}_{s}^{(k-1)})^{2}}{2\sigma_{2}^{2}})  &{\textrm{else}}
\end{cases},
\label{tau}
\end{equation}
where $\sigma_{1}$ and $\sigma_{2}$ are the parameters which are used to control the step of adjusting $\tau$. The larger value these two parameters take, the more gentle $\tau$ varies. By Eq. (\ref{tau}), the value of $\tau$ can be adaptively adjusted according to the dynamic threshold $\overline{u}_{s}^{(k-1)}$.


\begin{figure}
\centering
{\label{SS1}
\includegraphics[width=3.4in]{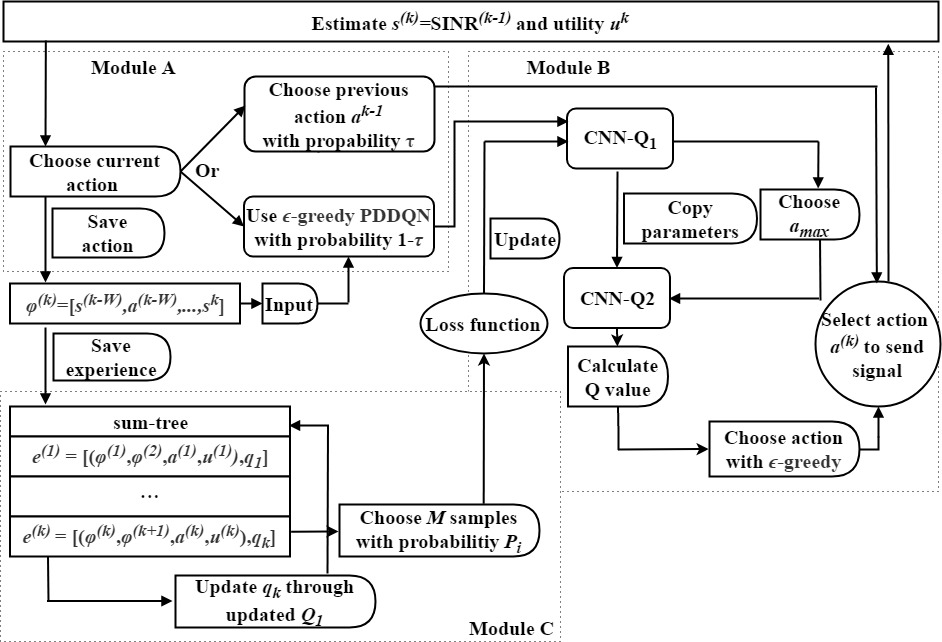}}
\caption{Diagram of the proposed algorithm.}
\end{figure}

\subsection{Double Deep Network Structure}
To update the Q value by Eq. (\ref{qvalue}) for each time slot, the neural networks can be used in our setup. As done in \cite{Han2017, Xiao2018a, Xiao2018b}, we use CNN with two convolutional layers (Conv) and two fully connected (FC) layers, too. The first and second Conv layers have 20 filters of size $3\times3$ and stride 1, and 40 filters of size $2\times2$ and stride 1, respectively. Rectified linear unit (ReLU) is used as the activation function in the two Conv layers. The first FC layer has 180 ReLUs, and the second FC layer has $S\times N$ outputs. Based on the outputs of CNN, the sender can obtain the optimal transmit power and channel.

Let $\bm{\varphi}^{(k)}$ denote the input of the CNN at time slot $k$, which consists of current system state and previous $W$ system state-action pairs, i.e., $\bm{\varphi}^{(k)} = [(s^{(k-W)},\textbf{a}^{(k-W)}),...,(s^{(k-1)},\textbf{a}^{(k-1)}),s^{(k)}]$. In this structure, we create two same CNNs which are denoted as $Q_{1}$ and $Q_{2}$, respectively. $Q_{1}$ is used to select the action $\textbf{a}_{max}^{(k)}$ that has the maximum Q value computed by Eq. (\ref{amax}), and $Q_{2}$ use $\textbf{a}_{max}^{(k)}$ to calculate the target Q value $Q_{target}$ by Eq. (\ref{qtarget}). The parameters of $Q_{1}$ and $Q_{2}$ at time slot $k$ are denoted as $\bm{\theta}^{(k)}_{1}$ and $\bm{\theta}^{(k)}_{2}$, respectively.

\begin{equation}
\textbf{a}_{max}^{(k)} = \textrm{arg}\max \limits_{\textbf{a}'\in \mathcal{A}}Q_{1}(\bm{\varphi}^{(k+1)},\textbf{a}';\bm{\theta}^{(k)}_{1}),
\label{amax}
\end{equation}

\begin{equation}
Q_{target}^{(k)}= u_{s}^{(k)} +\gamma Q_{2}(\bm{\varphi}^{(k+1)},\textbf{a}^{(k)}_{max};\bm{\theta}^{(k)}_{2}),
\label{qtarget}
\end{equation}
where $\bm{\varphi}^{(k+1)}$ denotes the next state. Note that in this letter the utility $u_{s}^{(k)}$ is considered as the reward, $\textbf{a}_{max}^{(k)}$ denotes an action that has the maximum Q value trained by $Q_{1}$ network at state $\bm{\varphi}^{(k+1)}$. The parameters of $Q_{2}$ are not updated, but overwritten by the parameters of $Q_{1}$ at a period of $f$ time slots.

\subsection{Prioritized Experience Replay}
Let $\textbf{e}^{(k)}=[\bm{\varphi}^{(k)},\bm{\varphi}^{(k+1)},\textbf{a}^{(k)},u_{s}^{(k)}]$ denote the experience sample of time slot $k$. And $\textbf{e}^{(k)}$ is stored in the sum-tree \cite{Schaul2015} (a kind of data structure where every node is the sum of its children and each leaf is associated with its own priority, denoted as $q_{k}$). Thus the whole sum-tree at time slot $k$ can be denoted as $\mathcal{ST}= \{(\textbf{e}^{(1)},q_{1}),...,(\textbf{e}^{(k)},q_{k})\}$. For each experience replay, we extract the first $M$ experience samples in order of descending probability from the sum-tree. For the sake of brevity, we use the index $i$ to represent the $i$th sample of these $M$ experience samples. Thus, we have that $\{(\textbf{e}^{(i)},q_{i})\}=\{([\bm{\varphi}^{(i)},\bm{\varphi}^{(i+1)},\textbf{a}^{(i)},u_{s}^{(i)}],q_{i})\}, 1\leq i \leq M$. The probability of the $i$th sample (denoted as $P_{i}$) is calculated by
\begin{equation}
P_{i} = \frac{q_{i}}{\sum_{j=1}^{k}q_{j}}.
\label{SS_E}
\end{equation}
To evaluate the priority of the experience sample, the temporal-difference (TD) error ($\psi _{i}$) is used, which is computed by
\begin{equation}
\psi _{i}=Q_{target}^{(i)}-Q_{1}(\bm{\varphi}^{(i)},\textbf{a}^{(i)}).
\label{TD}
\end{equation}
According to the stochastic gradient descent (SGD) algorithm, the parameters $\bm{\theta}_{1}^{(k)}$ of $Q_{1}$ network are updated by means of minibatch updates and the loss function chosen by \cite{Schaul2015} is as follows.
\begin{equation}
L(\bm{\theta}_{1}^{(k)})=\frac{1}{M}\sum_{i=1}^{M}\omega _{i}\psi _{i}^{2},
\label{SS_E}
\end{equation}
where $\omega_{i}$ denotes the importance sampling weights and can be calculated by
\begin{equation}
\omega _{i} = \frac{(M\cdot P_{i})^{-\lambda }}{\max \limits_{1\leqslant j \leqslant  k}\omega _{j}},
\label{SS_E}
\end{equation}
where $\lambda$ is a factor which is used to control the amount of importance sampling. In particular, the cases of $\lambda=0, 1$ indicate no and full importance samplings, respectively.

After the parameters of $Q_{1}$ are updated, the TD error $\psi_{j}$ ($j\in \{1,...,k\}$) of the experience sample is recalculated via Eq. (\ref{TD}) and the priority of each experience is updated by $q_{j}=|\psi _{j}|$. The parameters $\bm{\theta}_{2}^{(k)}$ of $Q_{2}$ network do not need to be updated immediately, but replaced by the parameters $\bm{\theta}_{1}^{(k)}$ of network $Q_{1}$ with frequency $f$. The overall algorithm process is shown in Algorithm 1.


\begin{algorithm}
\caption{$(\tau, \epsilon)$-PDDQN anti-jamming scheme.}
\KwOut{Transmission power and channel}
Initialization: $N$, $h_{s}$, $h_{j}$, $\beta$, $C_{m}$, $C_{s}$, $\gamma$, $T$, $\tau$, $\sigma_{1}$, $\sigma_{2}$, $\bm{\theta}_{1}$, $\bm{\theta}_{2}$, $W$, $M$, $q$, $f$, $\lambda=0.4$, $\epsilon=0.3$, $s^{(0)}=\textrm{SINR}^{(0)}$, $\mathcal{ST}=\varnothing$\\
\For{$k$=1,2,...}{

	 \If{$k\leq W$}
        {
            Choose channel $x_{i}^{(k)}$ and power $P_{s}^{(k)}$ randomly;\\
        }

      \Else
      {
            $z$ = random[0,1];\\

            \If{$z \leq \tau$}
            {
            	Let $x^{(k)}= x^{(k-1)}$, $p_{s}^{(k)} = p_{s}^{(k-1)}$;\\
            }

            \Else
            {
            	Input $\bm{\varphi}^{(k)}, \bm{\theta}^{(k)}$;\\
            	Obtain CNN's output $Q(\bm{\varphi}^{(k)};\bm{\theta}^{(k)})$\;
            	Choose $\textbf{a}^{(k)}=[x^{(k)},p_{s}^{(k)}]$ by $\epsilon$-greedy;
            }

            Receive $s^{(k+1)} = \textrm{SINR}^{(k)}$;\\
            Evaluate $u^{(k)}_{s}$ via (2) and update $\tau$ via (5);\\
            Let $\bm{\varphi}^{(k+1)}=[(s^{(k-W+1)},\bm{a}^{(k-W+1)}),...,s^{(k+1)}$];\\
            Store $(\bm{\varphi}^{(k)},\bm{\varphi}^{(k+1)},\bm{a}^{(k)},u_{s}^{(k)})$ in $\mathcal{ST}$ as $\textbf{e}^{(k)}$;\\

            \For{$i=1,2,...,M$}
            {
            Choose $\textbf{e}^{(i)}$ with probability $P_{i}$ via (8);\\
            Obtain $\textbf{a}_{max}$ via (6) and $Q_{target}$ via (7);\\
            Calculate $\psi_{i}$ and $\omega_{i}$ via (9) and (11) respectively;\\
       }

            Update $\bm{\theta}^{(k)}_{1}$ via (10);\\
            \For{i=1,2,...,k}
            {
            	Calculate $\psi_{i}$ for all experiences in $\mathcal{ST}$ via (9) and the updated network $Q_{1}$;\\
                Update $q_{i}$ in  $\mathcal{ST}$ by $q_{i}=|\psi_{i}|$;\\
            }
            }
            \If{$k$\%$f == 1$}
            {
            	Update parameters of $Q_{2}:\bm{\theta}^{(k)}_{2}=\bm{\theta}^{(k)}_{1}$;
            }

}
\end{algorithm}

\section{EXPERIMENTAL RESULTS}
In this section, a number of experiments is conducted to evaluate the proposed method. In our experimental setup, the parameters are set as $J=2$, $N=32$, $P_{S}=[1, 5, 10]$W, $P_{J} = [0, 4, 8, 10]$W, $h_{s}=h_{j}=0.5$, $\beta=1$, $C_{m}=1$, $C_{s}=0.2$, $\gamma=0.6$, $T=5$, $\sigma_{1}=0.8$, $\sigma_{2}=85$, $W=32$, $M=10$, $f=10$, and epoch=400.

First, we implement the existing value-based RL methods including QL, DQN, DDQN, and PDDQN in a wireless communication system. These four methods are all based on $\epsilon$-greedy. The SINR performance and convergence rate are analyzed and the results are shown in Fig. 3(a). As expected, PDDQN performs the best among the four methods. Hence, PDDQN is selected. This is the first time to our knowledge that PDDQN is applied to the anti-jamming wireless communications. Next, we compare the SINR performance between the variable and constant transmission power models under the same average power $\overline{P}$ condition. For a fair comparison, both of the models use PDDQN and the comparison results are given after the convergence. According to our experiment, PDDQN becomes completely convergent after 200 time slots, and the average power $\overline{P}$ is 6.19W in the variable power model. We can see from Fig. 3(b) that the variable power model obtains much higher SINR than the constant power model under the condition of the same average power $\overline{P}=6.19$W. Therefore, the variable transmission power model is adopted. Finally, we apply the proposed $(\tau, \epsilon)$-greedy to the four value-based RL methods, which are named $(\tau, \epsilon)$-QL, $(\tau, \epsilon)$-DQN, $(\tau, \epsilon)$-DDQN, and $(\tau, \epsilon)$-PDDQN. The curves of SINR performance and convergence rate are drawn in Fig. 4. It is clear that the convergence rate with the proposed $(\tau, \epsilon)$-greedy is greatly improved for the four RL methods. Moreover, the communication performance (measured by SINR) with $(\tau, \epsilon)$-greedy is also much better than those without $(\tau, \epsilon)$-greedy. This is mainly due to the fact the proposed algorithm can not only select the more valuable action in most cases but also seek for the most valuable action at a faster speed.

\begin{figure}[htbp]
\vspace{-0.3cm}
\setlength{\belowcaptionskip}{-0.5cm}
\centering
\subfigure[]{
\includegraphics[width=4.3cm]{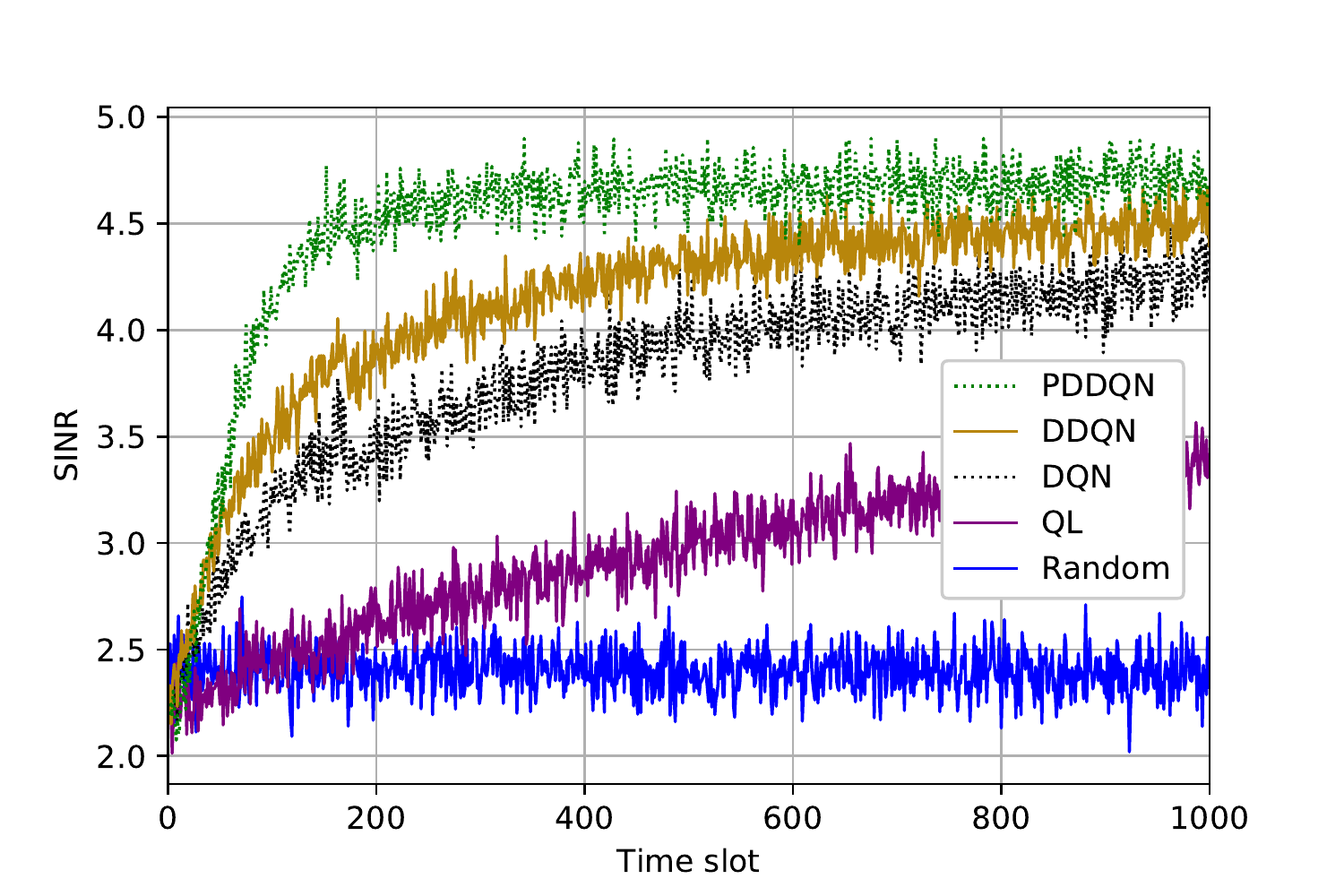}
}\hspace{-6.98mm}
\quad
\subfigure[]{
\includegraphics[width=4.3cm]{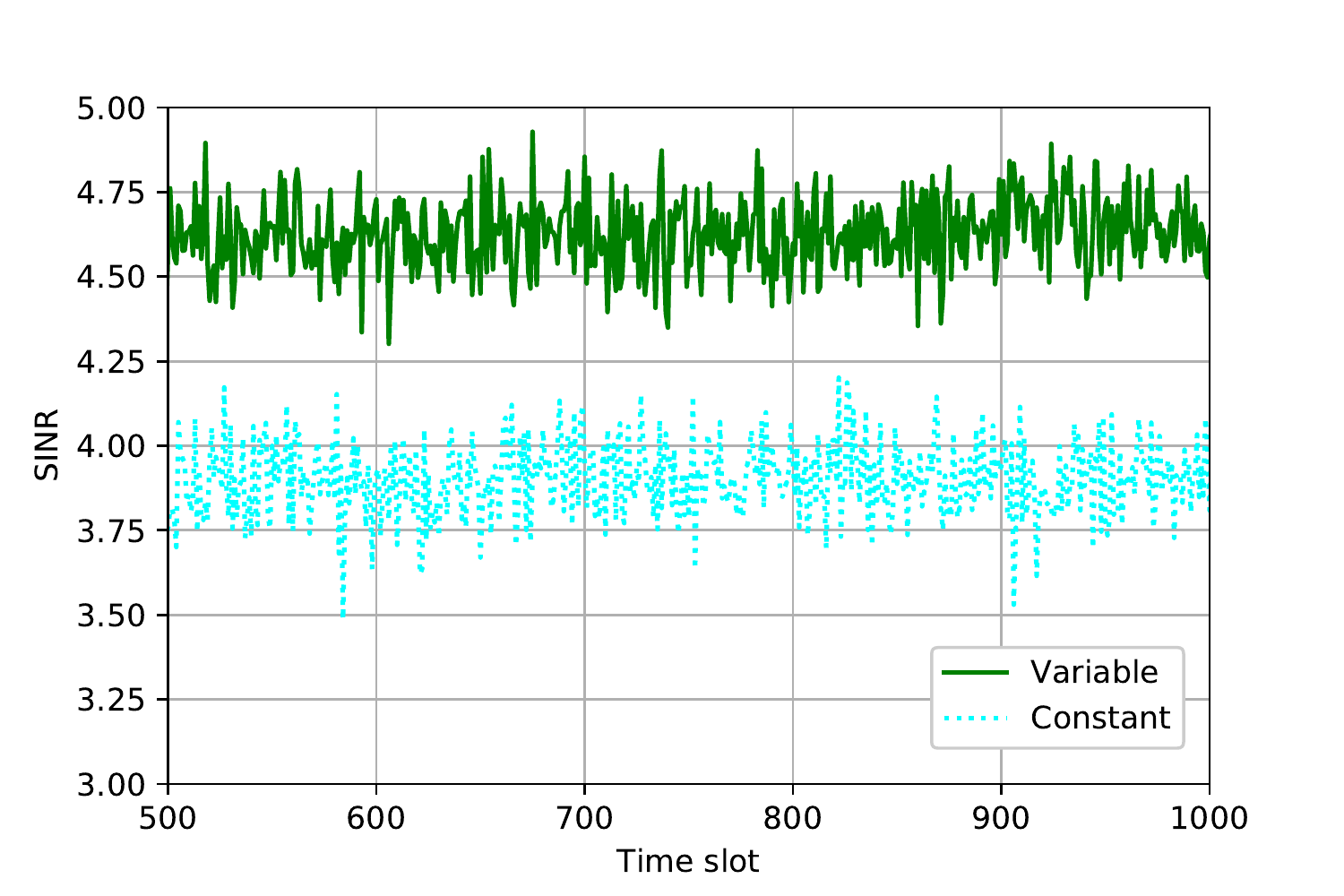}
}
\caption{Illustration of model selection. (a) SINR and convergence rate comparisons of various RL methods. (b) SINR comparison of the variable and constant power models.}
\end{figure}

\begin{figure}[htbp]
\vspace{-0.3cm}
\setlength{\belowcaptionskip}{-0.5cm}
\centering
\subfigure[]{
\includegraphics[width=4.3cm]{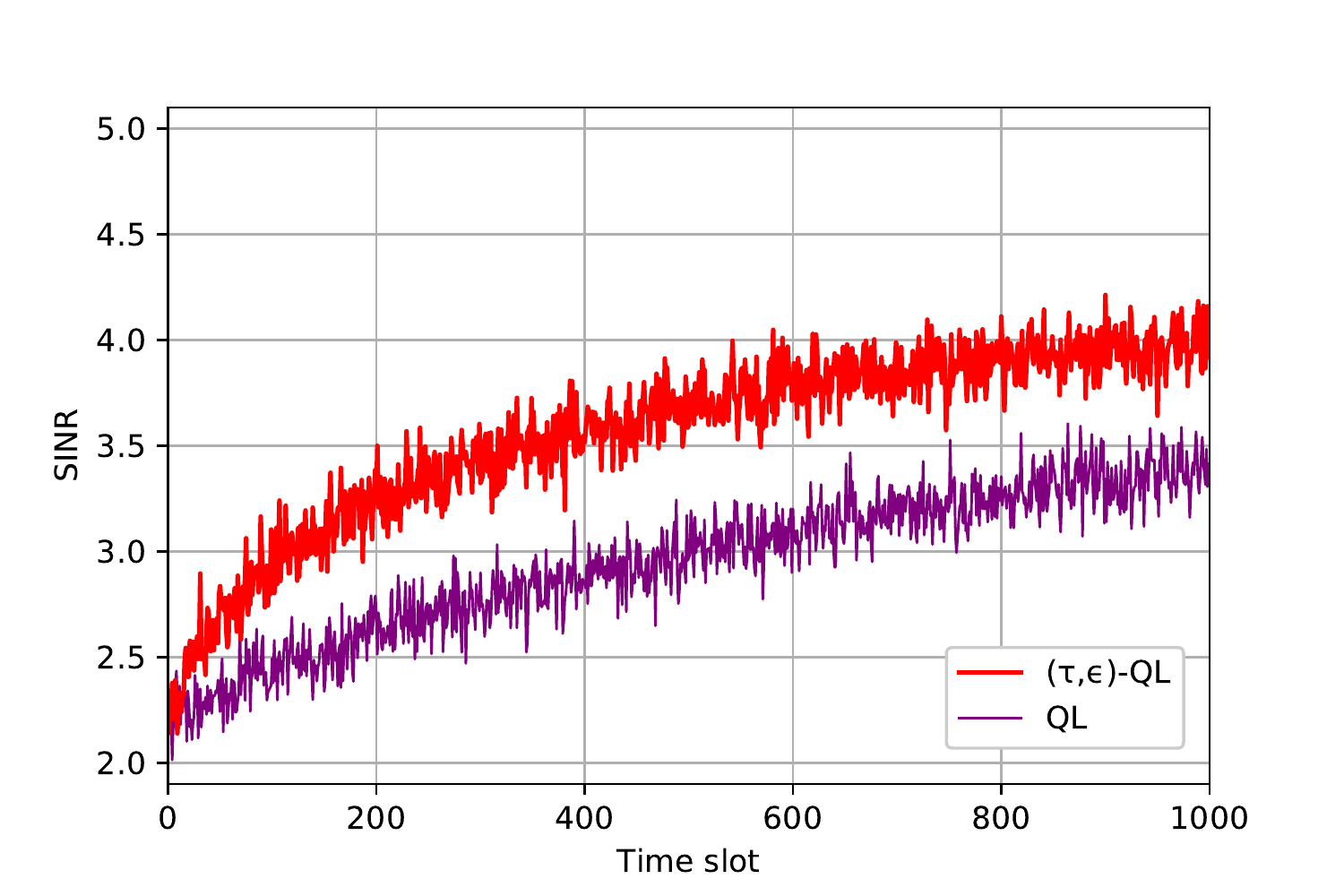}
}\hspace{-7mm}
\quad
\subfigure[]{
\includegraphics[width=4.3cm]{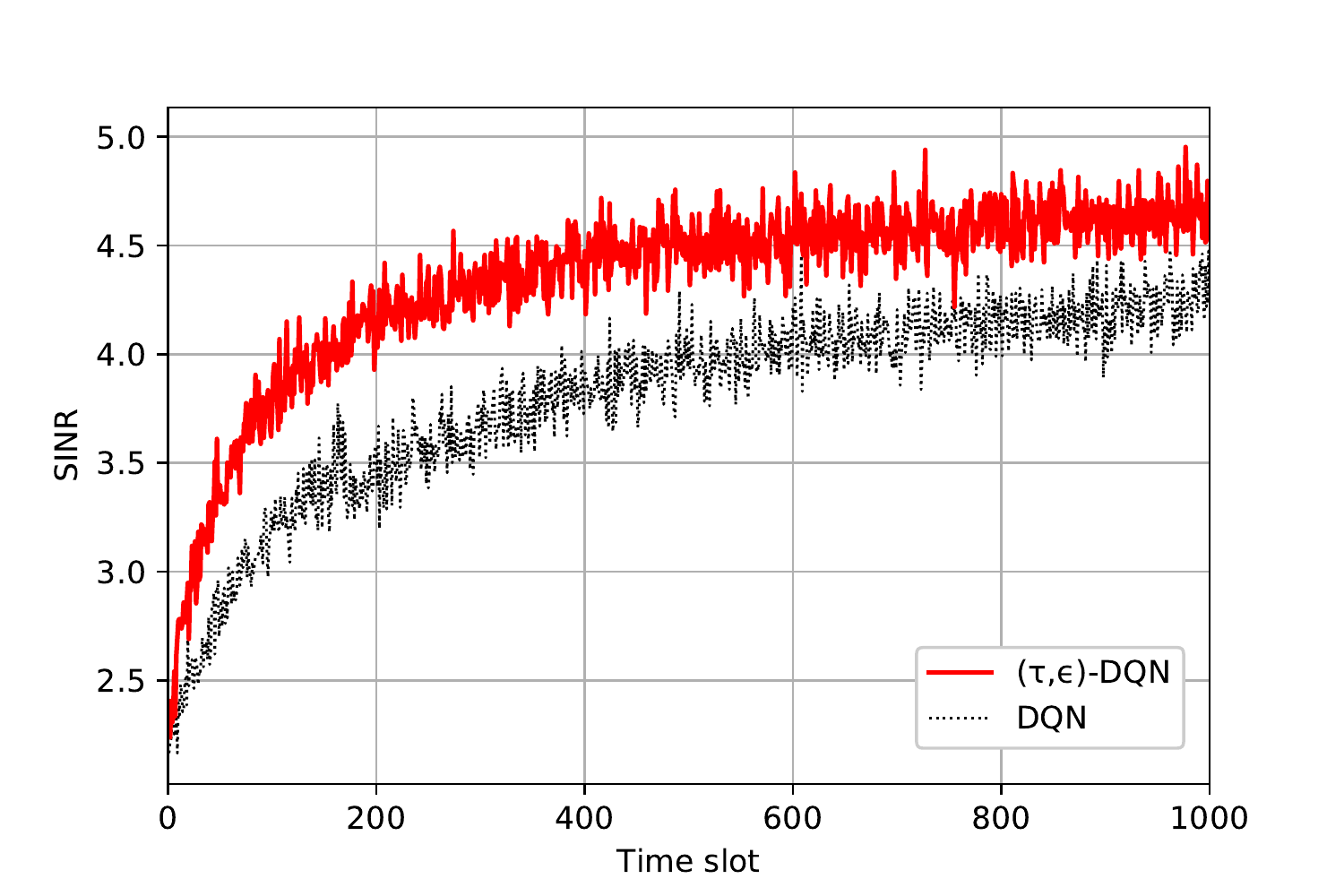}
}\hspace{-7mm}
\quad
\subfigure[]{
\includegraphics[width=4.3cm]{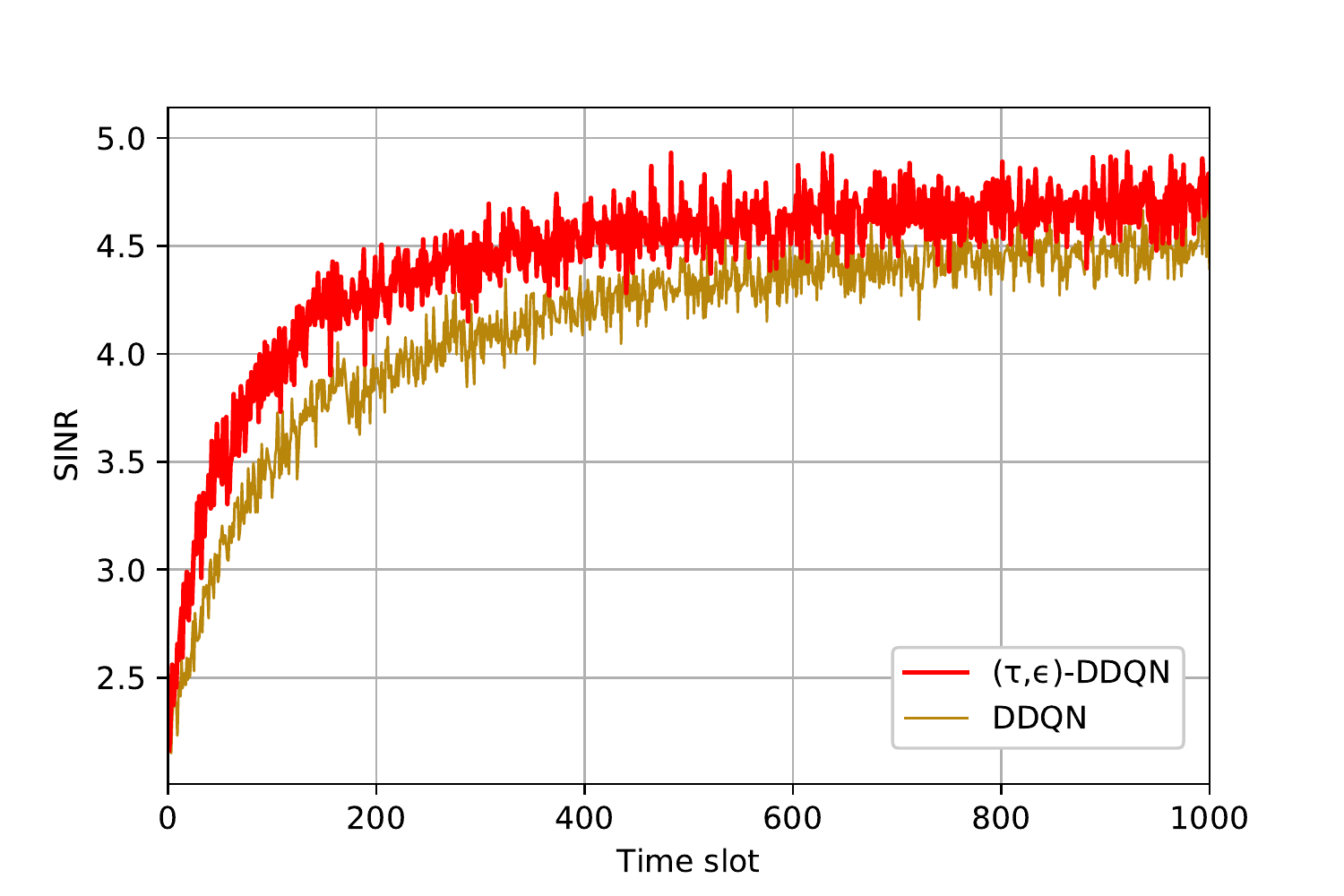}
}\hspace{-7mm}
\quad
\subfigure[]{
\includegraphics[width=4.3cm]{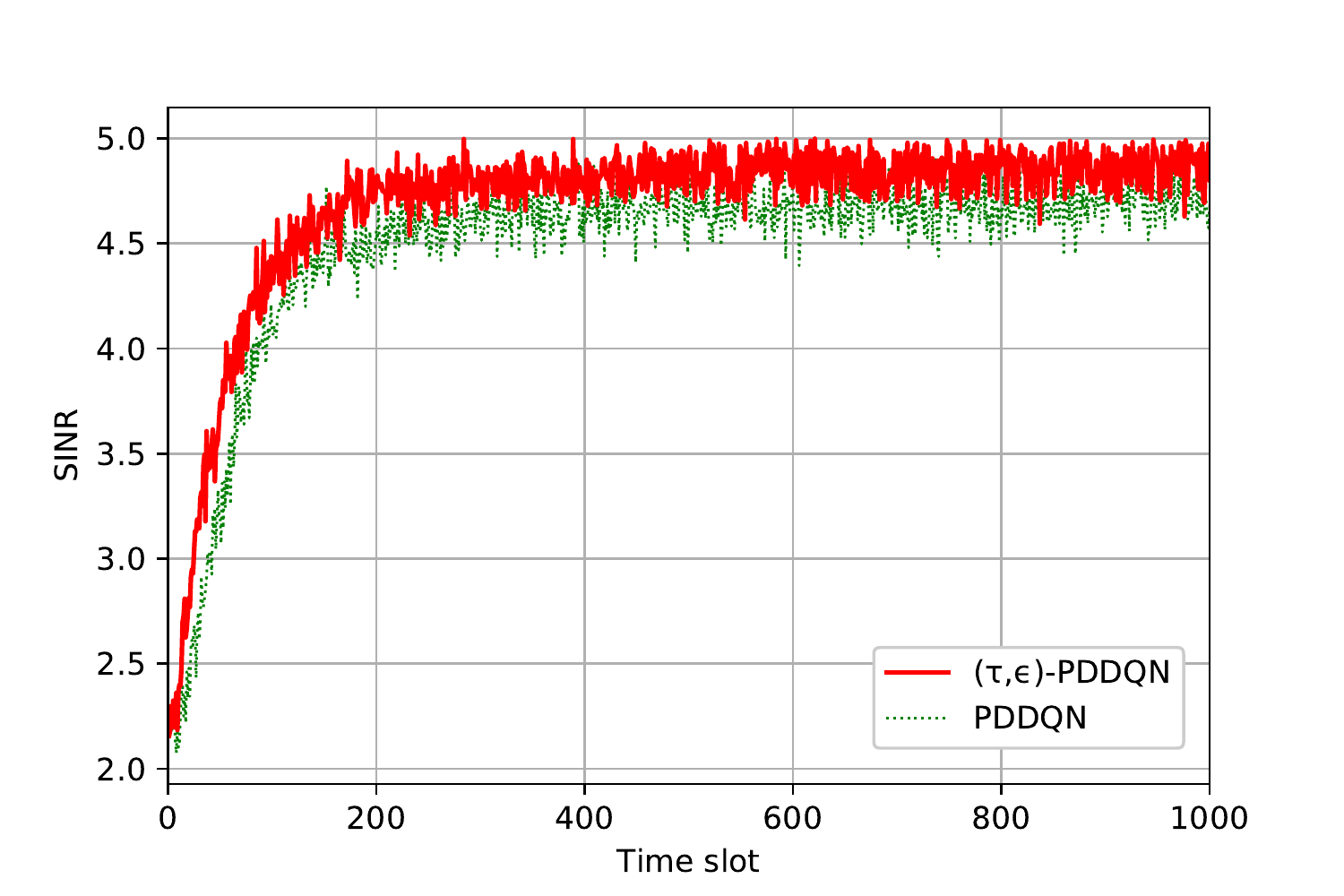}
}\hspace{-1.5mm}
\quad
\caption{SINR and convergence rate comparisons of various RLs with and without $(\tau, \epsilon)$-greedy.}
\end{figure}

\section{CONCLUSIONS}
In this letter, we have presented a $(\tau, \epsilon)$-greedy RL algorithm. Our proposed $(\tau, \epsilon)$-greedy can replace the existing $\epsilon$-greedy which is used in almost all the value-based RL methods. We have implemented the proposed algorithm in a wireless communication system against multiple jammers.  The experimental results show that the proposed algorithm accelerates the learning speed of the wireless device and significantly improves the performance of existing RL methods in terms of SINR and convergence rate. In the future, this algorithm will be implemented in the anti-jamming wireless communication scenario for continuous action set.


\bibliographystyle{IEEEtran} \small 
\bibliography{Manuscript}
\end{document}